# Conception d'un outil d'aide à l'indexation de ressources pédagogiques

## Extraction automatique des thématiques et des mots-clefs de documents UNIT

**Carlo Abi Chahine, Jean-Philippe Kotowicz, Nathalie Chaignaud, Jean-Pierre Pécuchet**

*INSA de Rouen - LITIS EA 4108*
*Place Emile Blondel - BP 8*
*76131 Mont-Saint-Aignan Cedex*
*{carlo.abi-chahine,kotowicz,chaignaud,pecuchet}@insa-rouen.fr*

RÉSUMÉ Le Learning Object Metadata (LOM) est le standard incontestable pour l'indexation des ressources pédagogiques. L'indexation de ces ressources en LOM est souvent accomplie manuellement par des documentalistes. Renseigner l'ensemble des champs du LOM d'un document est une tâche difficile et longue, nécessitant habituellement une lecture complète et une connaissance dans le domaine traité. Pour assister le documentaliste, un outil d'aide à l'indexation est nécessaire.

Afin de renseigner semi-automatiquement les champs du LOM, une analyse textuelle et structurelle de la ressource doit être accomplie. Dans cet article nous présentons un modèle fondé sur l'analyse textuelle de la ressource, permettant de suggérer aux documentalistes un ensemble de thématiques et de mots-clefs associés à ces thématiques. Pour cela, notre modèle prend en entrée une ressource pédagogique et une base de connaissances. Nous évaluons notre modèle sur un ensemble de documents UNIT en utilisant les liens de catégories Wikipedia comme base de connaissances.

MOTS-CLÉS : LOM, Indexation automatique et semi-automatique, extraction de mots-clefs

**1. Introduction**

L'indexation des documents pédagogiques est une tâche essentielle et complexe bien souvent accomplie à la main par les documentalistes. Avant d'extraire les mots-clefs d'un document, il est important d'identifier et de connaître les thèmes traités dans ce document. Pour des documents spécialisés dans un domaine particulier, l'expérience du documentaliste joue un rôle capital. Indexer des documents traitant de différents domaines s'avère encore plus complexe.

Dans le cadre du projet régional AICoTICE (Aide à l'Indexation et à la Cotation des ressources TICE), notre objectif est de concevoir, développer et évaluer un outil d'aide à l'indexation permettant aux documentalistes de l'« Université Numériques Ingénierie et Technologie » (UNIT)[1] de renseigner les notices électroniques correspondant à des documents pédagogiques. L'ensemble de ces documents forme un corpus hautement hétérogène en terme de structure (mise en page, format, etc.), de type de document (qcm, cours, examen, etc.) et de sujet traité (mathématiques, physique, économie, etc.). Le standard d'indexation utilisé par UNIT est le Learning Object Metadata (LOM)[2] qui permet de décrire exhaustivement les documents pédagogiques. Vu le nombre élevé de champs à renseigner, l'indexation est coûteuse en temps. De plus, les documents traitent de différents domaines et la variation interpersonnelle intervient beaucoup dans l'interprétation des documents et donc dans le choix des mots-clefs.

L'indexation automatique ou semi-automatique de ressources pédagogiques en format LOM nécessite plusieurs types de savoir-faire. L'extraction du contenant peut se faire par des méthodes d'analyse de structure de document associées à des méthodes d'apprentissage permettant d'extraire le titre, l'auteur et les différentes sections d'un document. En revanche, les méthodes de Traitement Automatique des Langues (TAL), de catégorisation ou de partitionnement permettent d'extraire les thématiques et/ou les mots-clefs à partir du contenu.

Cet article ne traite que de l'extraction des thématiques et des mots-clefs d'un texte. Nous proposons une méthode pour des outils d'aide à l'indexation. Le modèle prend en entrée une Ressource Termino-Ontologique (RTO) et un document textuel puis génère, en sortie, les thématiques et les mots-clefs contextualisés du document. La section 2 de cet article dresse un bref état de l'art du domaine de l'indexation manuelle des ressources pédagogiques et de l'indexation automatique. La section 3 décrit notre méthode d'extraction des thématiques et des mots-clefs contextualisés qui utilise un graphe représentant le contenu textuel des documents pédagogiques. La dernière section présente une évaluation de notre système qui utilise les liens des catégories de *Wikipedia*[3] comme RTO sur un corpus de documents UNIT.

---

[1] UNIT : http://unit.eu
[2] LOM : http://ltsc.ieee.org/wg12/20020612-Final-LOM-Draft.html
[3] Wikipedia : http://wikipedia.org

## 2. Etat de l'art

*2.1. Indexation des ressources pédagogiques*

[DE LA PASSARDIÈRE & JARRAUD 2005] explique les enjeux cruciaux de l'indexation des ressources pédagogiques d'un institut pour l'interopérabilité vis-à-vis des index d'autres instituts. L'étude internationale menée par l'ISO-CN36 a pour vocation d'analyser les contenus des index LOM de cinq instituts et conclut que souvent seuls quelques champs sont renseignés (titre, auteur, mots-clefs libres, etc.). Aussi, lorsque d'autres champs sont renseignés l'interopérabilité est mise à mal par des entorses aux contraintes que constituent les vocabulaires du LOM. Finalement, une des préconisations issues de l'étude est l'homogénéité de l'indexation, c'est-à-dire donner un sens à son index par rapport aux autres index.

Contrairement à C@mpusSciences, l'entrepôt UNIT n'impose ni de charte graphique à suivre, ni de métadonnées à utiliser. Or, pour mettre en place les préconisations de l'ISO-CN36, la lecture complète des ressources et leur compréhension deviennent indispensables pour une bonne indexation aux niveaux atomique et global. Cependant, pour UNIT, les documentalistes ne peuvent se permettre un tel luxe car le temps de traitement d'un document serait bien trop long.

Avant tout, le LOM a pour vocation première l'interopérabilité des index. Malgré cette volonté, les contenus de ses champs libres (mots-clefs, classification, etc.) sont trop hétérogènes (d'un point de vue qualitatif et quantitatif) selon les documents indexés et les personnes les indexant (variation interpersonnelle). L'idéal serait donc de proposer aux documentalistes des choix issus d'un vocabulaire contrôlé (dictionnaire, thésaurus, ontologie ou folksonomie [HUYNH KIM BANG & DANÉ 2008]). Des outils d'aide à l'indexation manuelle, comme Metalab [DE LA PASSARDIÈRE & JARRAUD 2004], pourraient être étendus pour suggérer et non imposer aux documentalistes une liste de termes pour les champs libres.

*2.2. Catégorisation et partitionnement des documents*

La catégorisation de texte consiste à affecter à un document une étiquette provenant d'un vocabulaire contrôlé en utilisant un corpus de documents étiquetés. Les méthodes d'apprentissage, telles que les classifieurs bayesiens [ZHENG 1998] ou les machines à vecteur support [LI et al. 2005], permettent de catégoriser des textes. Le partitionnement, quant à lui, divise le corpus de documents en grappes en ignorant tout vocabulaire contrôlé ou document étiqueté. Les documents d'une même grappe sont jugés « similaires ». La principale difficulté réside dans le choix d'un espace métrique et d'une distance associée acceptable (distance euclidienne, de Jacard, etc.) pour générer de bonnes grappes.

Dans le cadre des documents pédagogiques UNIT, il est difficile d'envisager d'utiliser ces méthodes pour l'extraction des thématiques. D'une part, le niveau actuel d'indexation des documents UNIT (servant de base d'apprentissage) est trop hétérogène et d'autre part, la mise en place d'une métrique s'avère difficile (due à la grande diversité structurelle et textuelle des documents).

*2.3. Indexation de documents en TAL*

En TAL, les méthodes telles que le TF-IDF (*Term Frequency-Inverse Document Frequency*) [SALTON 1983] et l'analyse sémantique latente [DEERWESTER et al. 1990] sont fréquemment utilisées pour extraire des termes significatifs à partir d'un corpus. L'idée de ces méthodes dites discriminantes, est qu'un terme est d'autant plus important qu'il est fréquent dans un texte d'un corpus et peu présent dans les autres. Cet aspect discriminatoire vis-à-vis des autres documents n'est pas envisageable dans notre contexte. En effet, dans un ensemble de documents pédagogiques traitant du même sujet, les mêmes termes sont utilisés à maintes reprises et donc jugés non-discriminants par ces méthodes alors qu'ils pourraient être des mots-clefs dans un contexte d'indexation, où chaque document doit être indexé indépendamment des autres. Aussi, ces méthodes fournissent des mots-clefs non-contextualisés car l'aspect sémantique des termes est ignoré.

Notre approche s'inscrit dans le type de famille d'indexation sémantique et conceptuelle. [BAZIZ 2005] discrimine l'indexation sémantique et l'indexation conceptuelle. D'une part, l'indexation sémantique consiste à lever l'ambiguïté existante dans le couple terme/sens. Cette activité selon [SANDERSON 1997] pourrait améliorer faiblement la précision d'un système de recherche d'information dans la mesure où la désambiguïsation est exacte. D'autre part, l'indexation conceptuelle s'appuie non seulement sur l'indexation sémantique, mais plus généralement sur l'utilisation de RTO comme charnière de l'outil d'indexation. [KHAN 2000] propose une méthode d'indexation des documents prenant en considération les cooccurrences des termes ainsi que la proximité sémantique de ces dernières par rapport à une RTO.

Dans l'approche présentée, nous souhaitons unifier cooccurrences et proximité sémantique en concevant et évaluant un algorithme paramétrable de fusion des graphes représentant les termes d'un document. La proximité sémantique sera déduite à partir d'une RTO. Dans un premier temps et à des fins expérimentales, nous avons conçu un modèle dont la RTO de support a une faible expressivité sémantique. Nous nous limitons ainsi à des relations entre concepts de type relation hiérarchique (hyponymie, méronymie), comme décrit dans [AIT EL MEKKI & NAZARENKO 2003] pour la construction d'index de fin de livre.

### 3. Représenter un texte par un graphe via une RTO

*3.1. Graphe représentant une entité du texte*

A l'instar de [KHAN 2000], notre méthode consiste à utiliser une RTO pour générer un graphe pondéré et orienté représentant chaque mot (appelé « entité ») et séquence de *n* mots (appelé groupe de *n* mots) d'un document . Nous comparons chaque entité du texte avec les entrées de la RTO (comparaison de chaînes de caractères ou d'expressions régulières). Si la comparaison est positive, le système génère un graphe dont le mot est une feuille (nœud terminal) en relation hiérarchique avec des nœuds concepts de la RTO. Chaque nœud est ensuite mis en relation hiérarchique avec d'autres nœuds concepts. A ce stage, chaque entité du texte est représentée par un graphe non pondéré. Ensuite, l'affectation des poids de chaque relation permet de définir son importance dans le texte. Un poids est affecté à chaque relation d'un graphe en fonction de son niveau de profondeur. Par exemple, si les concepts intéressants pour l'indexation se situent dans les niveaux inférieurs de la RTO, les poids des relations inférieures doivent être plus importants. La Figure 1 présente à gauche le graphe associé à l'entité « Souris » dans lequel nous donnons un poids supérieur aux relations les plus spécifiques (ici « Informatique » →« Souris »). Une profondeur maximale des graphes doit être fixée à l'avance afin d'éliminer des concepts trop génériques ou une dissémination sémantique trop importante par rapport à l'entité de départ.

### 3.2. Algorithme de fusion des graphes entités

La fusion des graphes construits à partir des entités d'un texte a pour but d'extraire des contextes et des thèmes abordés dans ce texte. Plus exhaustivement abordée dans [ABI CHAHINE et al. 2008], l'opération de fusion permet de fusionner deux graphes entités en renforçant les relations communes. Ce renforcement est paramétrable (par ex, addition des poids, addition pondérée des poids, etc.). L'opération considère également les relations exclusives aux deux graphes. L'utilisateur peut, par exemple atténuer les poids de relation n'existant que dans A et que dans B, mais également laisser les poids inchangés. La Figure 1 montre la fusion entre les graphes entités de « Souris » et « Clavier » en additionnant les poids des relations qu'ils ont en commun et en conservant les poids d'origine des autres relations.

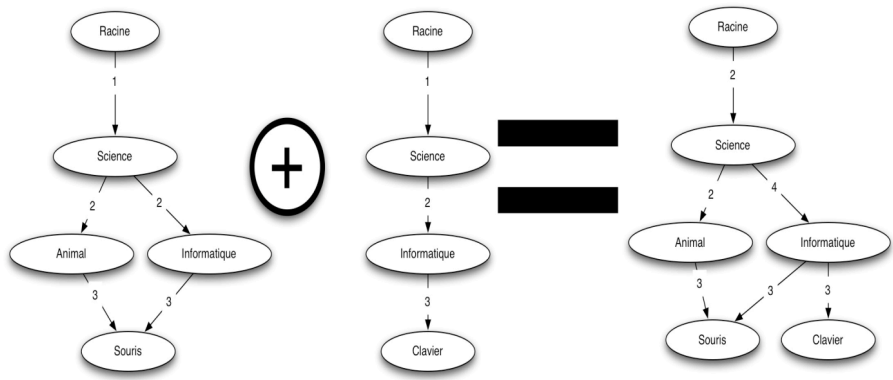

**Figure 1.** *Opération de fusion sur les graphes entités « Souris » et « Clavier »*

*3.3. Sélection des thématiques et des mots-clefs contextualisés*

Une heuristique permettant d'extraire les thématiques du graphe consolidé, consiste à sélectionner les nœuds ayant le plus grand flot sortant (somme des arcs sortants). Dans l'exemple décrit Figure 2, « Informatique » et « Science » ont un flot sortant de 6. Nous sélectionnons uniquement « Informatique » car il est à un niveau plus profond dans le graphe et est donc le plus représentatif (vis-à-vis du paramétrage de l'exemple). L'extraction des mots-clefs revient à extraire les nœuds dont les distances avec les nœuds thématiques sélectionnés sont les plus grandes. Dans l'exemple, « Souris » et « Clavier » sont à une distance égale à 3 d'« Informatique » et sont donc les mots-clefs de la thématique « Informatique ».

**4. Application, paramétrage et évaluation : UNIT et *Wikipedia***

Nous proposons d'appliquer cet algoritme aux documents PDF de UNIT en utilisant *Wikipedia* en tant que RTO. *Wikipedia* est une encyclopédie collaborative dont les contributeurs veillent à l'intégrité des contenus ; nous considérons les contributeurs comme expert de leur domaine. Les articles *Wikipedia* sont hiérarchisés en catégories et sous-catégories (appartenant à la folksonomie définie par les utilisateurs) que nous allons utiliser comme structure de RTO.

Afin d'évaluer notre système d'extraction des thématiques et des mots-clefs associés, nous devons prendre en compte les caractéristiques de UNIT. Les documents de UNIT ayant une vocation pédagogique, nous traitons en priorité les groupes de *n* mots qui portent un aspect sémantique plus précis [SMEATON 1992] (par ex, « Théorème de X », « Algorithme de X », etc.). Ainsi, s'il existe des nœuds communs entre le graphe représentant tous les groupes de n mots et un graphe de mot simple, nous fusionnons les deux graphes sinon nous ignorons le graphe de mot simple.

Le paramétrage étant fixé empiriquement (profondeur 5 ; pondération décroissante du niveau générique vers le niveau spécifique ; addition pour les poids des relations communes ; conservation des poids des autre relations), nous avons réalisé une évaluation subjective *a posteriori* sur l'ensemble des documents au format PDF de UNIT, c'est-à-dire juger si les thématiques et mots-clefs associés extraits par le système pouvaient servir à une indexation semi-automatique des documents. Le Tableau 1 résume l'évaluation du système en terme de rappel (proportion de « bons » mots-clefs ou thématiques trouvés) et de précision (pertinence des mots-clefs et thématiques trouvés) pour 56 documents au format PDF. Nos résultats sont meilleurs que ceux obtenus par d'autres méthodes d'indexation.

|  | Thématiques | Mots-clefs |
|---|---|---|
| Rappel | 76% | 85% |
| Précision | 70% | 85% |

**Tableau 1.** *Résultats de l'évaluation*

A titre d'exemple, le Tableau 2 montre le résultat d'une indexation avec notre outil mise en comparaison avec l'indexation manuelle UNIT sur un document traitant des instruments de mesure[4] (pour l'architecture et le génie civil).

| *Thématiques UNIT* | *Thématiques trouvées* | |
|---|---|---|
| • Géomatique, topographie<br>• Traitement signal et image<br>• Génie civil, génie urbain, aménagement<br>• Images fixes : analyse et traitement d'images | • Instrument_scientifique<br>• Métrologie<br>• Outil_de_mesure<br>• Instrument_de_mesure<br>• Grandeur_physique<br>• Mesures_en_géométrie | |
| *Mots-clefs UNIT* | *Mots-clefs trouvés* | |
| • Tourillonement<br>• Collimation<br>• Mesures<br>• Angles<br>• Topographie<br>• Théodolite | • Instruments_de_mesure<br>• Théodolite<br>• Distance_zénithale<br>• Angle<br>• Instrument_de_mesure_du_temps | • Inclinomètre<br>• Tachéomètre<br>• Distance_et_longueur<br>• Angle_plan |

**Tableau 2.** *Comparatif de l'indexation manuelle UNIT et de celle de notre outil*

Les thématiques trouvées dans *Wikipedia* nous semblent bien adaptées à l'indexation des documents malgré les différences avec celle fournie par l'expert UNIT. Notamment, notre modèle ne retourne pas « Topographie » car c'est une notion liée à la « Géographie » dans *Wikipedia*. Concernant les mots-clefs, « Collimation » et « Topographie » sont liés à la thématique « Topographie » et ne sont pas des mots-clefs selon notre outil. Quant à « Tourillonement », il n'existe pas dans *Wikipedia*. Cependant, nous retournons des mots-clefs pertinents non fournis par UNIT tels que « Distance Zénithale », « Inclinomètre » ou « Tachéomètre ».

## 5. Conclusion et perspectives

Notre modèle semble innovant mais a encore besoin d'améliorations. En effet, en exploitant uniquement les informations textuelles d'un document, nous obtenons un rappel et une précision de bon niveau pour l'extraction des thématiques et des mots-clefs contextualisés d'un document. Nos perspectives pour améliorer ces résultats est d'intégrer la structure des documents pour renforcer (ou atténuer) certaines parties du contenu textuel. Par exemple, les titres affichés avec une police de grande taille et en caractère gras sont traités comme les autres parties du texte. Or, dans les documents pédagogiques, les titres sont porteurs d'une grande indication sur les thématiques et les mots-clefs.

De plus, dans le cadre de l'aide à l'indexation des notices LOM, nous souhaitons étudier la possibilité d'utiliser partiellement notre modèle, afin d'extraire les compétences nécessaires à la compréhension d'un document pédagogique mais aussi les compétences acquises après utilisation de cette ressource. En effet, l'indexation des thématiques et des mots-clefs est essentielle mais l'identification des compétences pré-requises et acquises donnerait encore plus de sens à l'indexation dans le contexte des sciences de l'éducation et de la formation [GRANDBASTIEN & HUYNH-KIM-BANG 2008].

---

[4] http://www.univ-valenciennes.fr/coursenligne/topographie/partie2/papier.pdf

## 6. Bibliographie